# Imaging of microscopic features of charged-particle tracks in a low-pressure gas

V.Dangendorf, H.Schuhmacher, U. Titt[*], K. Tittelmeier
Physikalisch-Technische Bundesanstalt, 38116 Braunschweig, Germany
* present addr.: Massachusetts General Hospital, Boston MA 02114, USA

**Abstract:**
An imaging system for measuring the spatial distribtion of charged particle tracks in a low-pressure gas is presented. The method is based on an optically read out time projection chamber. Results of experiments with fast heavy ions are shown.

**1. Introduction**
Radiation action in tissue depends on the microscopic details of the energy transfer in the nanometre range. Since it is not possible to measure with such a high resolution in condensed phase, various methods have been developed to substitute these measurements a) by measurements in low-pressure gases or b) by simulating the radiation transport through matter by Monte Carlo (MC) models and calculating the deposited energy or the number of ionisation events in the volumes of interest.

Most of the experimental approaches utilise tissue equivalent proportional counters, (TEPCs) whose geometrical size, scaled by the density ratio between tissue and gas simulates the size of the biological entity of interest. These methods are restricted to detector sizes which correspond to tissue volumes of a few 100 nm. On the other hand, the sizes of the radiation-sensitive cell structures range from several 100 nm down to 2 nm. An adequate instrument should cover the full range of these sizes.

In recent years an imaging system was developed at PTB which is able to measure the spatial pattern of energy deposition in a simulated cavity a few micrometre in diameter with a potential resolution of about 40 nm. With this instrument the full statistical correlation of ionisation events along the track of charged particles in the volume range from a few 10 nm to several micrometres is obtained. Furthermore, it is aimed as benchmark for testing results of MC calculations.

**2. Experimental method and results**
The experimental method is based on a time projection chamber with a parallel drift field, charge and proportional scintillation stages and optical readout (Optical PArticle Chamber, OPAC). Fig 1 shows the detector principle. A description can be found in [1]. The chamber is operated with triethylamine (TEA) vapour at a pressure of typically 1 kPa. In the past this measurement system was successfully employed in light ion beams at the accelerator facilities at PTB and Frankfurt University [1-3].

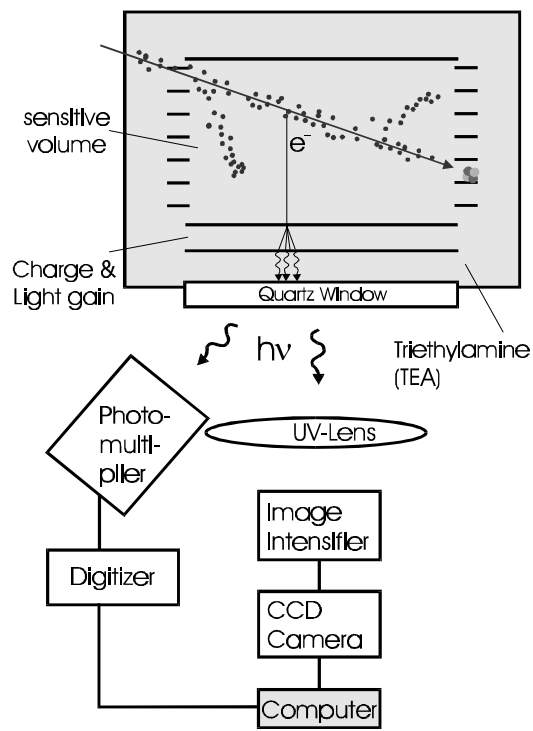

Figure 1: Schematic view of the OPAC

Recently measurements were performed at GSI / Darmstadt to demonstrate the feasibility of measuring tracks of fast heavy ions. It was shown, that the imaging system can handle the very large dynamic range in ionisation density between core and penumbra of such tracks, ranging from more than $10^3$ electrons per $mm^3$ gas down to few electron clusters or even single electrons in the sparsely ionising δ-electron tracks escaping from the core. In a first experiment 11,2 MeV/u Ar and 200 MeV/u C-ions were measured at the accelerator facilities of GSI. Fig. 2 shows a measured track of each kind of ion.

The energy of the δ-electrons produced by fast ions like the 200 MeV/u C-ions can be of the order of several hundred keV. To stop these electrons in the interaction volume of the OPAC the pressure required in the detector would to be several hundred kPa. Operating the chamber at such a high pressure would degrade the spatial resolution for the simulated sites to more than 1 micrometre. However, these high energy electrons are quite rare. The most interesting part of the track is the core and the part of the penumbra made up by the more frequent electrons of up to a few keV energy.

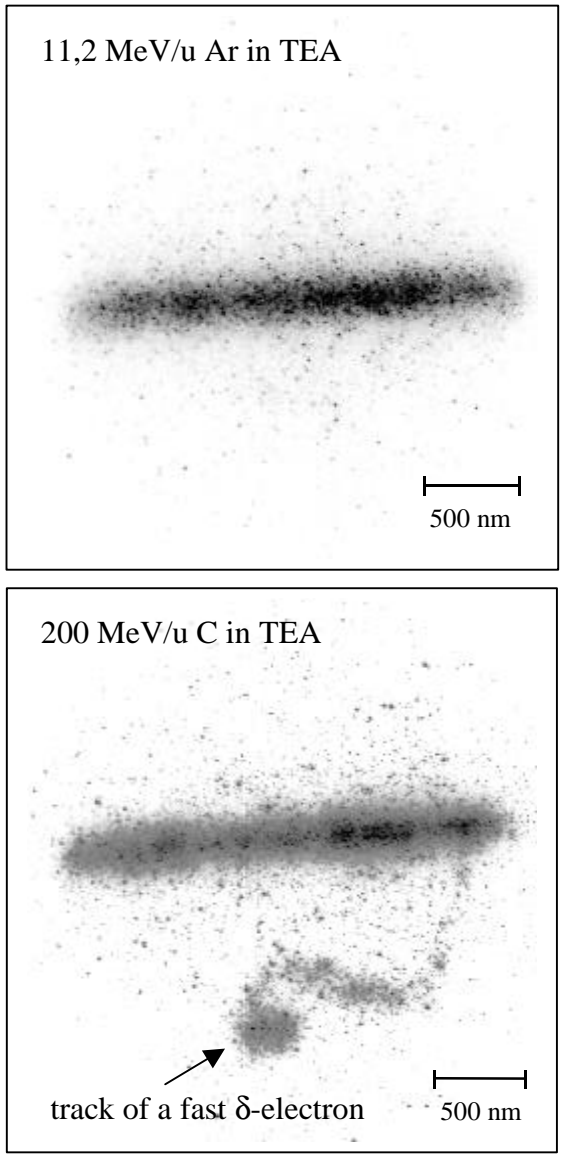

Fig.2: Examples of measured ion tracks in 1 kPa TEA. The specified length is scaled to tissue density

By varying the gas pressure in the chamber the track core with its high ionisation density can be investigated with a high spatial resolution (i.e. low gas pressure) and the penumbra with its long ranging δ-electrons with a lower resolution (i.e. at higher gas pressure). The stability of the chamber, irradiated with fast heavy ions (11,2 MeV/u Xe and 200 MeV/u C) was studied in a wide pressure range and found satisfactory in the range 0.5 – 4 kPa. Also operation at higher pressure will be studied in future by using Ar / TEA mixtures which according to [5] are expected to have good amplification and scintillation properties also at higher pressure.

### 3. Outlook

Based on this experience we will measure the spatial ionisation distribution of a variety of ion beams. In detail we will investigate C-ions in TEA gas and TEA based gas mixtures in the whole energy range relevant for heavy ion therapy. From these data the radial and longitudinal ionisation distribution will be obtained which will be compared to simulated track structure data obtained from the models of Krämer et al. [4].

Another property of fast ion tracks to be studied is based on the fact, that particles of same LET but different charge vary enormously in their speed. This results in a different δ-electron energy distribution which influences the transversal ionisation density in the particle track. In [2] this effect is used to identify light ions like protons and alpha particles by measuring LET and the transversal width of the tracks. At GSI we will apply this technique also to heavier particles from carbon up to uranium and compare the results with MC-calculations [4].